\documentclass[final]{aa}
\usepackage{graphics,epsf,epsfig,graphicx}
\usepackage{natbib}
\usepackage{url}
\usepackage{psfrag}
\usepackage{amsmath,amssymb}
\usepackage{txfonts}
\usepackage{changebar}

\bibpunct{(}{)}{;}{a}{}{,} 

\begin{document}

\title{On the stratified dust distribution of the GG Tau circumbinary ring}

\author{C. Pinte\inst{1,2} \and L. Fouchet\inst{3,4} \and F. M\'enard\inst{1}
\and J.-F. Gonzalez\inst{3} \and G. Duch\^ene\inst{1}}

\offprints{C. Pinte}
\institute{Laboratoire d'Astrophysique de Grenoble, CNRS/UJF UMR~5571, 
414 rue de la Piscine, B.P. 53, F-38041 Grenoble C\'edex 9, France
\\e-mail: {\tt pinte@obs.ujf-grenoble.fr,
  menard@obs.ujf-grenoble.fr, duchene@obs.ujf-grenoble.fr}
\and
School of Physics, University of Exeter, Stocker Road, Exeter EX4 4QL, United Kingdom
\and 
Universit\'e de Lyon, Lyon, F-69003, France;
Universit\'e Lyon 1, Villeurbanne, F-69622, France;
CNRS, UMR 5574, Centre de Recherche Astrophysique de Lyon;
\'Ecole Normale Sup\'erieure de Lyon, 46 all\'ee d'Italie,
F-69364 Lyon cedex 07, France
\\e-mail: {\tt Laure.Barriere@ens-lyon.fr, Jean-Francois.Gonzalez@ens-lyon.fr}
\and
ETH Hoenggerberg Campus, Physics Department, HPF G4.2, CH-8093 Zurich, Switzerland}

\authorrunning{}
\titlerunning{}

\date{Received ... / Accepted ...}

\abstract{}{Our objective is to study the vertical dust
distribution in the circumbinary ring of the binary system GG Tau
and to search for evidence of stratification, one of the first steps
expected to occur during planet formation.}
{We present a simultaneous analysis of four scattered light images
spanning a range of wavelength from 800~nm to 3800~nm and compare them
with (i) a parametric prescription for the vertical dust
stratification, and (ii) with the results of SPH bi-fluid hydrodynamic
calculations.}
{The parametric prescription and hydrodynamical calculations of
stratification both reproduce the observed brightness profiles
well. These models also provide a correct match for the observed
 star/ring integrated flux ratio. Another
solution with a well-mixed, but ``exotic'', dust size distribution
also matches the brightness profile ratios but fails to match the
star/ring flux ratio. 
These results give support
to the presence of vertical stratification of the dust in the ring of
GG Tau and further predict the presence of a radial
stratification also.}{}

\maketitle

\section{Introduction}

The first step of planet formation is the coagulation of dust grains,
during low speed collisions \citep[][ and references
therein]{Dominik06PPV}. Micrometer size grains grow to form aggregates
that will finally give birth to kilometer size planetesimals and
eventually planets.  In parallel to grain growth, dust grains are
expected to settle towards the disk midplane due to the conjugate
actions of the stellar gravity and gas drag.

Dust settling arises because gas is pressure-supported (and thus has a
subkeplerian velocity) whereas dust is not. Therefore, gas slows down
the dust resulting in the dust settling towards the midplane and
spiraling inwards to the central star because of angular momentum
conservation \citep[see][]{Weidenschilling77,Stepinski96}. This
settling is highly dependent on the grain size with the larger grains
($>$~1~m) almost unaffected by gas drag and the smaller ones
($<$~1~$\mu$m) so strongly coupled to the gas that they follow its
motion. For the intermediate sizes, the dust distribution decouples
strongly from that of the gas \citep{Barriere05}.

The result is the formation of a dust sub-disk close to the equatorial
plane \citep[e.g.,~][]{Safronov69,Dubrulle95}, that may fragment and
produce planetesimals when the density of the dust layer exceeds the
gas one \citep{Goldreich73}. Despite remaining questions and
uncertainties -- how dust grains overcome the ``meter size barrier''
without being accreted onto the star \citep[e.g.\
][]{Weidenschilling77} or destroyed by high speed collisions
\citep{Jones96,Blum00}, how Kelvin-Helmholtz instabilities prevent the
dust sublayer to fragment -- models predict that dust grains can grow
up to very large sizes and settle towards the midplane, leaving a
population of small grains close to the surface and larger grains
deeper in the disk.

Such a stratified structure should have consequences on disk
observables like their spectral energy distributions (SEDs) and
scattered light images \citep{Dullemond04}. So far however, most
studies of dust settling have been based on SED analysis only
\citep{Apai04,Furlan05}.

\defcitealias{Duchene04}{D04}
\def\D04{\citetalias{Duchene04}}
An alternative approach to probe the disk structure is through
high resolution scattered light images. Such images of protoplanetary
disks are becoming available over a wide wavelength range in a few
cases, from the optical with HST, to mid-IR with ground-based adaptive
optics.
Analysis of these images with model fitting were usually
restricted to {\sl a single} wavelength but multi-wavelength
  studies are becoming more and more common
(see \citealp{Watson06PPV} and references therein). Because
scattering properties are highly dependent on grain size and because
different wavelengths probe different depths into the disk, scattered
light images over a wide range of wavelengths are potentially a
powerful tool to detect dust settling.  Indeed, attempts to
simultaneously fit all wavelengths with a single model under the
assumption of well mixed gas and dust indicate that the grain size
distribution is likely more complex than that of the ISM
\citep{Mathis77}. In particular, a dependence of the maximum grain
size with vertical distance above the disk midplane (vertical
settling) sometimes appears to be needed \citep[][hereafter
D04]{Duchene04}.

GG Tau is of special interest to study dust settling. Its geometry,
well known from millimeter emission maps \citep{Guilloteau99}, is
simple: the dust is distributed in a circumbinary ring between 180 and
260\,AU and the inclination is known, around 40$^\circ$, allowing a
wide sampling of the scattering phase function. Multiple wavelength
studies of the disk (\D04) have shown that I-band scattered light
images probe sub-micron grains located about 50\,AU above the disk
midplane, whereas at longer wavelengths, in the L' band, observations
probe deeper regions of the disk (25\,AU above the disk midplane) and
reveal the presence of grains larger than 1\,$\mu$m, strongly
suggesting the presence of a stratified structure.

In this paper, we investigate (i) whether multiple wavelength
scattered light images of the GG Tau ring can be interpreted in terms
of a disk model with a stratified structure and (ii) whether such a
structure is required to explain the observations. In
Sect.~\ref{sec:modeles_GG}, we present model fitting of synthetic
scattered light images. Two families of models are explored: (i) models
where dust and gas are perfectly mixed in the disk and where optical
properties of the grains are varied and (ii) models
that include a stratified structure described in a parametric way.
In Sect.~\ref{sec:modeles_hydro}, we explore if the model with a
stratified structure, in better agreement with observations, is 
compatible with a more physical (hydrodynamical) description of dust spatial distribution.
The effect of  realistic dust settling on scattered light images of
the GG Tau ring is studied  by coupling hydrodynamical simulations
of the dust distribution with a radiative transfer code. We discuss
our results in
Sect.~\ref{sec:discussion} and conclusions are presented in Sect.~\ref{sec:conclusion}.

\section{Parametric models of the ring \label{sec:modeles_GG}}

\subsection{Model description}
\label{sec:modeles_GG_descr}

Model fitting is performed on the multiple wavelength scattered light
images obtained in the Hubble Space Telescope filters F814W, F110W,
F160W\footnote{In the rest of the paper, we will  refer to the
F814W, F110W, F160 filters as I, J, H bands for commodity.} by
\cite{McCabe02} and \cite{Krist02} and on Keck-AO images in L' by \D04.

Scattered light images are calculated by means of the Monte Carlo
radiative transfer code MCFOST \citep{Pinte06}. The general concept is
to follow individual photon packets that randomly scatter in the disk,
with probabilities given by the radiative transfer equations and
dust optical properties, track them, and produce images when they
exit the model boundaries.

In this paper, we focus our analysis on the stratified structure of
the ring and we adopt the same ring geometry and dust grain
composition as \D04. The disk is assumed to be in hydrostatic
equilibrium with a Gaussian vertical profile. The model parameters are
the inner and outer radii (190 and 260\,AU respectively), the total
dust mass (0.0013\,$M_\odot$, \citealp{Guilloteau99}), the vertical
scale height (21\,AU at 180\,AU), the flaring index $H(r) \propto
r^{1.05}$ and the surface density exponent $\Sigma (r) \propto
r^{-1.7}$. The density distribution extends inside the inner radius
with a radial Gaussian decrease with a $1/2e$ width of 2\,AU~:
\begin{equation}
  \Sigma(r) = \Sigma(190\,\textrm{AU}) \exp
  \left(-\frac{1}{2}\,\left(\frac{r-190\,\textrm{AU}}{\Delta r_0}\right)^2\right)
\end{equation}
with $\Delta r_0 =2$\,AU. The density is set to zero inside
180\,AU.

We assume dust grains to be homogeneous spheres (Mie theory) and adopt
the porous dust grain optical properties from \citet[model
A]{Mathis89}, with an 0.5\,g.cm$^{-3}$ average grain density. The
grain size distribution $n(a,\vec{r})$ is defined locally, in each
cell of the computation grid, and may then depend on the position
within the disk, allowing to model disks with a stratified
structure. Integrated over the whole disk, the grain size distribution
is described by a power-law, $\mathrm{d} n(a) \propto a^{p} \mathrm{d}
a$. The minimum grain size is taken to be $a_\mathrm{min}=0.03\,\mu$m
and $a_\mathrm{max}$ and $p$ are considered as free parameters. 
With $a_\mathrm{max}=0.9\,\mu$m and $p=-3.7$, this dust model
reproduces the interstellar extinction law up to 8\,$\mu$m
\citep{Mathis89}.  \D04 showed that neither a grain distribution
typical of interstellar medium nor the distribution used by
\cite{Wood02} to reproduce the SED of HH~30 are able to account
simultaneously for both the scattered light images and I-band
polarisation rates. In this section, we try to construct a grain size
distribution that would reproduce both observed quantities
simultaneously

Models are monochromatic and are calculated at the central wavelength
of the different filters (0.81, 1.0, 1.6 and 3.8~$\mu$m for the F814W,
F110W, F160W and L' filters).  Each model is computed with a total number of
128 million photon packets. The pixel size is 0.04\,\arcsec and images
are convolved with a Gaussian kernel with a resolution similar to
those of observations, i.e.,  FWHM$= 0.065\,$\arcsec,
$0.080\,$\arcsec, $0.128\,$\arcsec\ and $0.091$\,\arcsec for I, J, H
and L' bands, respectively.

The quality of the models was determined by a $\chi^2$ minimization
method.  The disk geometry was validated in previous studies
\citep[e.g.,\ ][]{Guilloteau99,Silber00,Duchene04} and in this paper
we focus instead on the scattering properties, in particular the
scattering phase function. In a first step, we only consider the azimuthal
brightness profiles in our fitting procedure. These profiles were
constructed with the same method as described in \D04 (see their
section 4.2) using $10^\circ$ sectors. For each wavelength, we define
a reduced $\chi^2$, based on the uncertainties in the images. Model
uncertainties are negligible with respect to the observations due to
the large number of photon packets generated for each model.  We
finally define a total $\chi^2$ as the sum of the $\chi^2$ at each
wavelength. From this total $\chi^2$, a Bayesian analysis is performed
to estimate the probability of occurrence of each parameter value. It
also gives an estimate of the validity range of each ``best-fit''
value (see for instance \citealp{Press92} and also
\citealp{Lay97} for an application to millimeter observations of
HL~Tau). Because our model is simplistic (the observations
  reveal systematic departures from the axysymmetry assumed in the model), this $\chi^2$ should not be
 interpreted in statistical terms but as defining a metric to compare
 the general behaviour of different models\footnote{ The statistical
   interpretation of the Bayesian analysis remains however valid
   because systematic deviations between observations and models add a
 constant shift to the $\chi^2$ that does not modify the probability distributions.}.

\subsection{Models without stratification}
\label{sec:modeles_GG_no_strat}
Studies of grain growth processes show that grains with fractal
dimension and/or with a size distribution that departs from a
power-law, and whose scattering properties are probably noticeably
different from those of previously cited distributions, are produced.

Modification of the grain size distribution or use of complex grain
size distributions results in a dramatic increase of the parameter
space.  Here, our aim is rather to verify how one can, by increasing in
a reasonable manner the parameter space, modify the grain distribution
used by \D04 so that it reproduces simultaneously all observations,
without invoking any stratification.

Scattered light images at multiple wavelengths probe different grains
inside a single distribution. Thus, for a grain size distribution
typical of the ISM with $a_\mathrm{max} = 0.9\,\mu$m, the median
scatterer has a size of 0.5\,$\mu$m in the I band and 0.7\,$\mu$m in
the L' band (\D04).

A simple way to play on the characteristic size observed at each
wavelength is to modify the relative fraction of each grain size via
the slope of the size distribution. The goal is to introduce large
grains, susceptible to reproduce the L' band image, and therefore
larger than 1\,$\mu$m, but in a small enough number so that their
contribution at small wavelengths is negligible.

To do so, we explore a 2 dimensional parameter space, by varying the
maximum grain size between 1\,$\mu$m (size slightly smaller than the
size required to reproduce the L' band image in the work of \D04) and
100\,$\mu$m (size large enough that the contribution from these
grains is negligible on scattered light, even at 3.8\,$\mu$m), and the
index of the grain size distribution between $-3.5$ (as in the ISM) and $-6.0$, by
steps of 0.5.

\begin{figure}[t]
  \centering
   \psfrag{Angle}[c][c]{Position angle ($^\circ$)}
  \psfrag{Intensite}[c][c]{Normalized intensity}
  \includegraphics[width=\hsize]{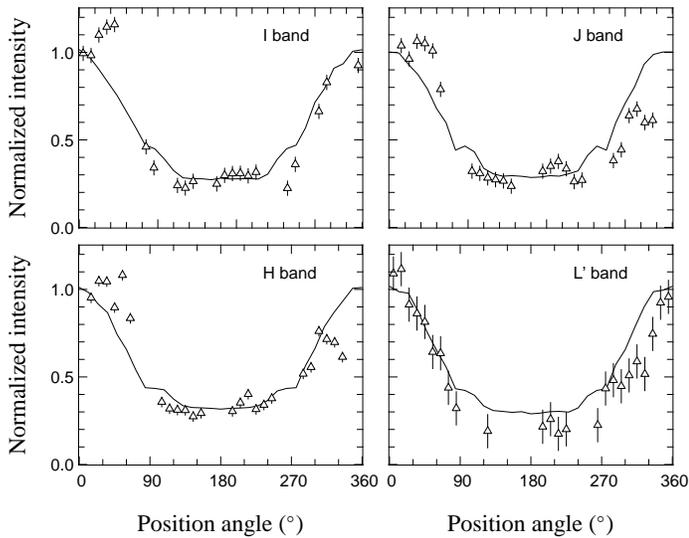}
  \caption{Azimuthal brightness profiles of the best model without
    stratification (solid line) superimposed to observed profiles
    (triangles). The model parameters are $p = -5.5$ and
    $a_\textrm{max} = 100\,\mu$m.
     The position angle $0^\circ$ corresponds
    to the disk semi-minor axis, on the front face of the ring. \label{fig:fit_gg_no_strat}}  
\end{figure}

Figure~\ref{fig:fit_gg_no_strat} presents azimuthal brightness profiles
for the best model. Results are in agreement with observations at all
wavelengths with a global $\chi^2$ of 57.28
(Table~\ref{tab:chi2_no_strat_gg}). 
The best model reproduces very well the images in the I, J and H bands but slightly
overestimates the flux on the back side of the ring in the L' band.

\begin{table}[b]
  \centering
  \caption{$\chi^2$ of the best models without stratification.
    Each line gives the $\chi^2$ in the I, J, H and L' bands, as well as the
    total $\chi^2$ defined as the sum of the individual ones
    for the models minimizing one of these $\chi^2$, respectively.
 \label{tab:chi2_no_strat_gg}}
 \begin{tabular}{lrrrrrrr}
 \hline
 \hline
 Model  & $\chi^2_I$ & $\chi^2_J$ & $\chi^2_H$ & $\chi^2_L$ & $\chi^2$ \\
 \hline
Best model I band & \bf{11.81} & 16.70 & 31.75 & 25.77& 86.03 \\ 
Best model J band & 12.56 & \bf{16.35} & 25.56 & 3.47& 57.94 \\ 
Best model H band & 12.84 & 19.23 & \bf{23.95} & 21.45& 77.47 \\ 
Best model L' band& 19.66 & 26.45 & 55.62 & \bf{1.72}& 103.45 \\ 
 \hline
Best model  & \bf{12.16}& \bf{16.69}& \bf{25.99}& \bf{2.43}& 57.28\\
 \hline
 \end{tabular}
\end{table}

\begin{figure}[t]
  \centering
 \psfrag{proba}[c][c]{Probability}
  \psfrag{p}[c][c]{$p$}
  \psfrag{amax}[c][c]{$a_\mathrm{max}$}
  \includegraphics[width=\hsize]{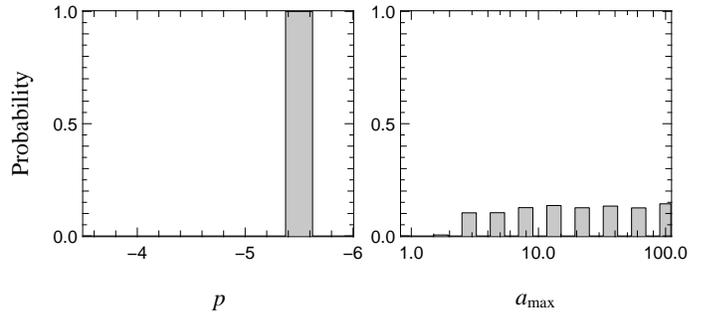}
 \caption{ Bayesian probability distribution in the non-stratified case
   for grain size distribution slope (left) and maximum grain size (right).
 \label{fig:proba_gg_no_strat}}
\end{figure}

The Bayesian analysis shows that a single index for the grain size
distribution, equal to -5.5, can reproduce the observations
(Fig.~\ref{fig:proba_gg_no_strat}).  While our crude sampling prevents
us from accurately defining the allowed range for the slope index, we
can readily exclude values smaller than -6.0 and larger than -5.0, leading to a tight
constraint. Regarding the maximum grain size, we obtain that a minimum value of
3\,$\mu$m is required but all values beyond that are equiprobable.

\subsection{Models with stratification}
\label{sec:modeles_GG_strat}

We model the vertical stratification of the disk with a simple parametric
law where the vertical scale height of the grains depends on the grain
size:
\begin{equation}
  h_0(a) = h_0(a_\mathrm{min}) \,
  \left(\frac{a}{a_\mathrm{min}} \right)^{-\xi}\ .
\end{equation}
In the case of a perfect mixing between gas and dust, 
 $\xi=0$ and
$h_0(a) = h_0(a_\mathrm{min})$ is constant and corresponds to the gas
 scale height.

\begin{figure}[t]
  \centering
  \includegraphics[width=0.7\hsize,angle=190]{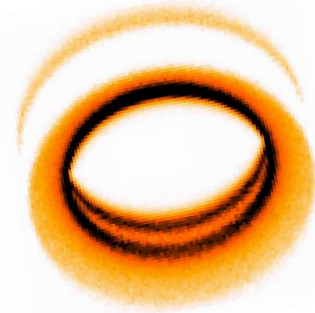}%
  \caption{Synthetic image of the GG Tau ring in the I band with only vertical
  stratification. The color map is in logarithmic scale and is inverted
    with bright regions in dark. The image size is 3.7 arcsec.\label{fig:GG_pb}}
\end{figure}

The particular geometry of the GG Tau ring in which the back side of
the inner rim of the ring is seen directly suggests that such a
stratified model is too simplistic to describe scattered light
observations. In Fig.~\ref{fig:GG_pb} a synthetic map obtained in the
I~band with the stratification described above is shown. The inner
edge of the ring, observed on the back face, presents a vertical
brightness gradient (the surface is about twice as bright as the
midplane) which is not seen in observations \citep{Krist02,Krist05}.

This gradient is caused by the fact that, as a function of altitude at
which photons hit the ring, they do not scatter on grains of the same
size. Photons that scatter in the upper layers of the ring interact
with small grains, which are isotropic scatterers and send a
significant fraction of energy back to the observer. On the contrary,
photons that scatter close to the midplane interact with larger
grains, which are more forward-scatterers and send only a smaller
fraction of energy back to the observer\footnote{Albedo effects are
superimposed on those of the phase function. The albedo also varies
vertically in the disk, but the relative differences are much smaller
than for the phase function.}.

\begin{figure}[t]
  \centering
\includegraphics[width=0.75\hsize]{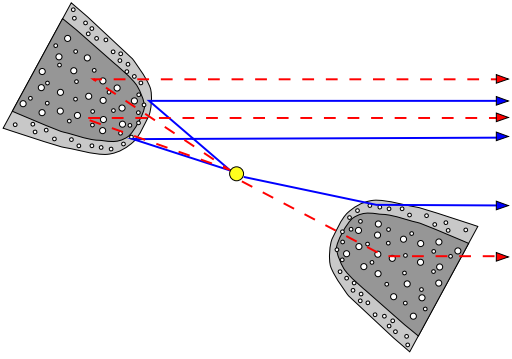}  
  \caption{Sketch of the vertical and radial stratification proposed
  to account for the scattered light images of GG
  Tau.\label{fig:strat_rad}}
\end{figure}

Thus, the different scattered light images cannot be explained
\emph{only} with vertical stratification. A complementary
stratification, in the radial direction appears needed to explain the
globally uniform brightness observed on the back side of the ring. A
configuration where small grains would form a layer enclosing larger
grains, as presented in Fig.~\ref{fig:strat_rad}, might produce images
in agreement with observations. At optical wavelengths, photons would 
scatter in the surface layer and meet small grains, whatever their height
relative to the disk midplane. In the infrared, photons would penetrate
deeper in the ring and scatter on larger grains.

To represent such a stratification, we introduce an analytical
description, similar to the one used for the vertical stratification,
so that the density remains a continuous function. The characteristic
length of the radial decrease of the surface density, $\Delta r$,
becomes a function of the grain size, that we choose to describe with
a power-law, whose index is the same as the vertical stratification one:
\begin{equation}
  \Delta r (a) = \Delta r_0 \left( \frac{a}{a_\mathrm{min}} \right)^{-\xi} 
\end{equation}
with $\Delta r_0$ fixed to 2\,AU.

We then explore a 2D parameter space, playing with maximum grain size
between 1 and 100\,$\mu$m  as previously and the
stratification index, $\xi$, that we sample between 0 (no
stratification) and 0.5.  In this section, we set the index $p$ of the
integrated grain size distribution to a constant value of -3.5,
following models from \D04.

\begin{figure}[t]
  \centering
  \psfrag{Angle}[c][c]{Position angle ($^\circ$)}
  \psfrag{Intensite}[c][c]{Normalized intensity}
  \includegraphics[width=\hsize]{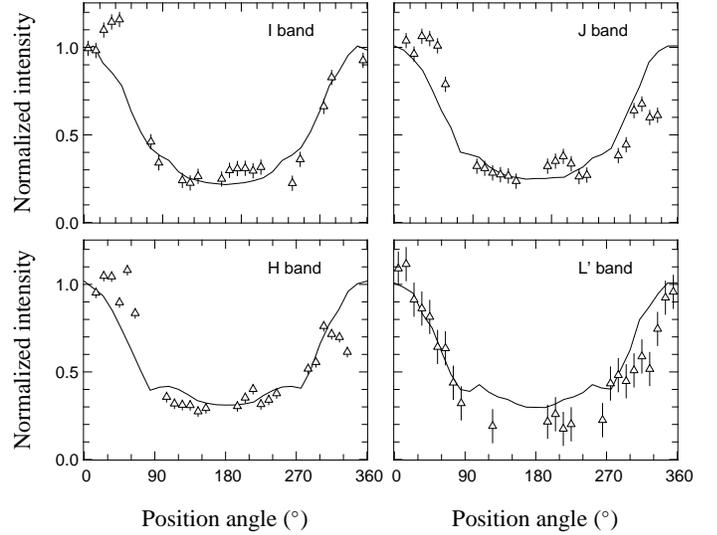}
  \caption{Azimuthal brightness profiles of the best model with stratification
    (solid line) superimposed to observed profiles (triangles). The
    model parameters are $\xi = 0.15$ and $a_\textrm{max} = 60\,\mu$m.
    \label{fig:fit_gg_strat}}
\end{figure}

Figure~\ref{fig:fit_gg_strat} presents azimuthal brightness profiles
obtained for the best model, compared to observed profiles.  Results
are in good agreement with observations at each wavelength, with an
azimuthal profile \emph{similar at all wavelengths}. The intensity on
the back side of the ring, in the L' band is slightly larger, by about
1~$\sigma$, with respect to the observations. In order to quantify
further the quality of the fit, we compare $\chi^2$ values obtained for
the best model to those that we get by fitting one single wavelength
(Table~\ref{tab:chi2_strat_gg}). The global $\chi^2$, 57.70, is
comparable to the one obtained from models without dust stratification
(57.28, see Table~\ref{tab:chi2_no_strat_gg}).

\begin{table}[b]
  \caption{$\chi^2$ of the best models with stratification. The different
    lines have the same meaning as in Table~\ref{tab:chi2_no_strat_gg}.
    \label{tab:chi2_strat_gg}}
  \centering
  \begin{tabular}{lrrrrrrr}
 \hline
 \hline
     Model  & $\chi^2_I$ & $\chi^2_J$ & $\chi^2_H$ & $\chi^2_{L'}$ & $\chi^2_\mathrm{tot}$ \\
 \hline
Best model I band  & \bf{9.10} & 16.67 & 30.11 & 2.44& 58.31 \\ 
Best model J band  & 10.17 & \bf{14.95} & 32.45 & 7.93& 65.50 \\ 
Best model H band  & 14.44 & 20.54 & \bf{25.11} & 32.13& 92.23 \\ 
Best model L' band & 17.09 & 21.82 & 48.20 & \bf{1.30}& 88.41 \\ 
 \hline
Best model  & 10.18& 16.19& 29.03& 2.30& \bf{57.70}\\
 \hline
  \end{tabular}
\end{table}

We note that, in the I, J and H bands, the best global model is only
marginally poorer that the best models at each of these wavelengths
but that the difference is more important for the L' band similarly to
the models without stratification.

\begin{figure}[t]
  \centering
  \psfrag{proba}[c][c]{Probability}
  \psfrag{strat}[c][c]{Stratification index $\xi$}
  \psfrag{amax}[c][c]{$a_\mathrm{max}$}
  \includegraphics[width=\hsize]{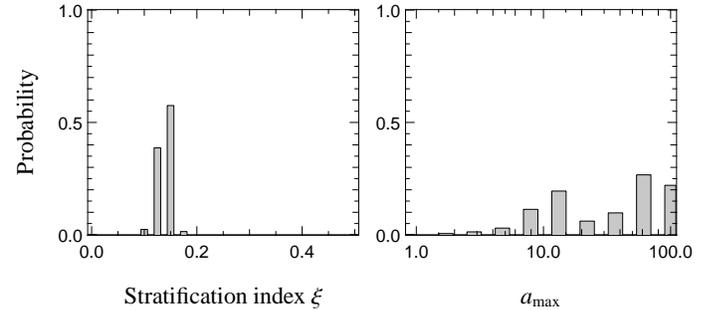}
  \caption{Bayesian probability distribution in the stratified case.
  \label{fig:proba_gg_strat}}
\end{figure}

Results of the Bayesian analysis are presented in
Fig.~\ref{fig:proba_gg_strat}. The stratification index is well
constrained, with a probability curve peaking between 0.125 and 0.15
and a dispersion of about 0.025. Regarding the maximum grain size, it
appears that values smaller than $5$\,$\mu$m are excluded but that
beyond this limit, all values are roughly equiprobable, grains larger
than $5$\,$\mu$m contributing very little to the scattered flux at
wavelengths smaller than 3.8\,$\mu$m.

\subsection{Polarisation and flux ratio}
\label{sec:modeles_GG_pola}

Both models, with radial and vertical stratification or with a slope for the size
distribution equal to $-5.5$, present equally good fits to the azimuthal
brightness profiles. In order to discriminate between these two
solutions, we compare their predictions of the polarisation and of the
disk-to-star flux ratio with observations.

\begin{figure}[t]
  \psfrag{pola}[c][c]{Linear polarisation level}
  \psfrag{angle}[c][c]{Position angle ($^\circ$)}
  \includegraphics[width=\hsize]{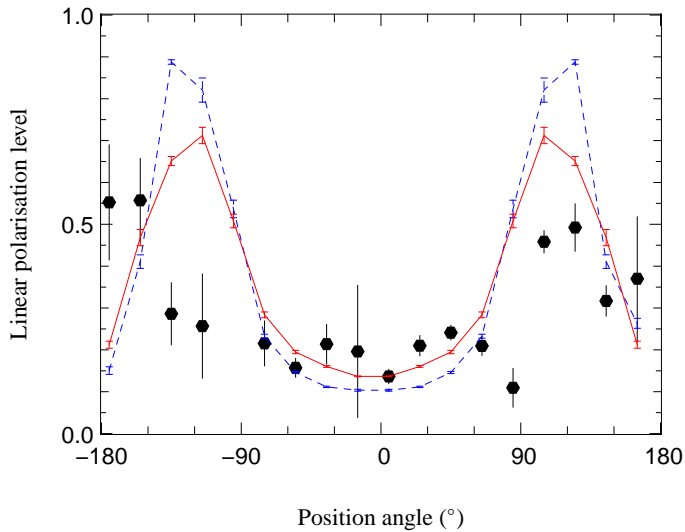}
  \caption{Observed  (from \citealp{Silber00})  and synthetic  polarisation at 1\,$\mu$m as a function
    of the azimuthal angle. Polarisation maps have been binned by a
    factor 4, resulting in a pixel size of 0.172'' and then divided in 20$^\circ$
    sectors where the polarisation is measured, for each sector, in the
    pixel of maximum brightness. Black circles
    represent the polarisation extracted from observations. The solid red line
    corresponds to the best model with stratification and a slope $p=-3.5$
    and the dashed blue line to the model without stratification and
    a slope $p=-5.5$. The position angle $0^\circ$ is defined as in
    the previous figures and corresponds to the disk semi-minor axis,
    on the front side of the disk. \label{fig:pola_gg}}
\end{figure}

Figure~\ref{fig:pola_gg} presents the observed polarisation in the J
band \citep{Silber00} as a function of the azimuthal angle compared
with the polarisation predicted by the two best models and shows the
differences between them. The model with stratification presents a
polarisation rate around 15\,\% on the front side and peaking around
65-70\,\% on the back side of the disk whereas for the model without
stratification but with a grain size distribution slope of $-5.5$, the
polarisation rate is 10\,\% on the front side and a maximum
polarisation level of 80-90\,\% on the back side. None of the two
models provides a good fit to the observations of \cite{Silber00}. The
model with stratification performs slightly better ($\chi^2$ = 16 against a
$\chi^2$ of 33 for the model without stratification) and reproduces
reasonably well the polarisation in the regions where it is the best
defined, the front face of the disk.  The model without
stratification results in somewhat lower polarization rates (on the front side). 
However, both models overestimate the polarisation on the back
side of the disk. This behaviour is due to the high polarisability of the
porous grains used here. An exploration of the dust properties,
beyond the scope of the paper, is needed to obtain a better
agreement with observations and,  at this point,  polarisation cannot be
used as a strong 
constraint to discriminate between the two  models.

Table~\ref{tab:rap_flux_gg_R} gives all disk-to-star flux ratios for
observations and models. These measurements depend on the ring
geometry, in particular to its projected surface on the plane of the
sky. Because we did not explore this parameter, we focus here on the
variations of these ratios with wavelength.  The model with dust
stratification is in better agreement with observations: the flux
ratio is roughly constant between the I and H bands and decreases by
about 15\,\% in the L' band.  Even though it does not exactly
reproduce the observations where the ratio slightly increases from the
I to H bands, and then decreases in the L' band ($\approx 35\,\%$
smaller than in the I band), the global behaviour is respected, which
is not the case for the model without stratification where a steady
flux ratio decrease with wavelength is witnessed, with a factor around
6 between the I band and the L' band.

\begin{table}[b]
  \caption{Disk-to-star flux ratios as a function of wavelength.
  \label{tab:rap_flux_gg_R}}
  \begin{tabular}{llll}
 \hline
 \hline
Flux ratio & Observations & Models with    & Models without \\ 
(\%)       & \citep{Duchene04}  & stratification & stratification \\ 
\hline
I & $1.30 \pm 0.03$ & 0.92 & 0.57 \\
J & $1.40 \pm 0.03$ & 0.89 & 0.46 \\
H & $1.54 \pm 0.03$ & 0.91 & 0.30 \\
L'& $0.97 \pm 0.09$ & 0.77 & 0.10 \\ 
\hline
  \end{tabular}
\end{table}

Results presented in this section were based, for both cases, on
the best model according to $\chi^2$ minimizations. The study of
brightness profiles only gives a lower limit for the maximum grain
size. Polarisation and colours show the same insensitivity
to the precise value of $a_\mathrm{max}$ as the phase function.
Results presented are therefore valid for all values
$a_\mathrm{max} \gtrsim 5\,\mu$m and, in particular, cannot help to
better constrain this parameter.

\section{Hydrodynamical modeling}
\label{sec:modeles_hydro}

The parametric modeling of GG~Tau's circumbinary ring presented in the
previous section points towards a stratified disk
where both vertical and radial segregation of grains are needed.

In this section, we aim at testing the validity of this hypothesis,
using a more physical modeling of the processes of vertical settling and
radial migration of dust grains. Results of hydrodynamical
simulations, where dust settling and migration are computed more
realistically, are used as an input to the radiative transfer code to
produce synthetic scattered light images.

\subsection{Hydrodynamical code}
\label{sec:SPH}

To compute the hydrodynamical evolution of gas and dust, we use
Smoothed Particle Hydrodynamics (SPH), a Lagrangian technique
described by \citet{Monaghan92}. The equations and approximations are
rigorously established by \citet{Bicknell91}.

Our code was derived from Murray's code \citep{Murray96} and is
described in \citet{Barriere05}. It is fully 3D, bi-fluid, non
self-gravitating and locally isothermal with the temperature law given
by $T(r)\propto r^{-3/4}$, where $T$ is the temperature and $r$ the
distance to the star projected onto the midplane. The initial surface
density is given by $\Sigma \propto r^{-3/2}$, following estimations
of the Minimum Mass Solar Nebula (MMSN) and of \cite{Guilloteau99}.

The dust phase, representing one single grain size at a time,
is treated as a pressureless fluid and the coupling
between gas and dust is done through aerodynamic drag
\citep[see][]{Monaghan95,Monaghan97}. The drag exerted by dust on gas
is neglected \citep[see details in][]{Barriere05}.

In the initial state, dust and gas are considered as well mixed, which
means that the dust disk is flared. This is correct for the smaller
grain sizes when dust is strongly coupled to the gas, but for larger
grain sizes, we can expect grains to start close to the midplane
because they are produced by the growth of smaller grains that have
already settled. Still, up to 1~mm, grains settle quickly enough to
the midplane that this initial state does not really matter.  The
contribution of larger grains to the synthetic images is negligible
(see Sect.~\ref{sec:modeles_GG_no_strat}).

\subsection{Parameters}
\label{sec:param_hydro}

\begin{table}
\caption{Simulation parameters for the circumbinary disk}
\label{tab:paramCB}
\begin{tabular}{@{}l@{\ =\ }ll@{\ =\ }l@{}}
\hline\hline
$M_{\star1}$ & $0.5\ M_\odot$ &
$a$ & $ 35\ AU\qquad\qquad e\ =\ 0.25$ \\
$M_{\star2}$ & $0.65\ M_\odot$ &
$s$ & 10, 7.5, 5, 2.5, 1~$\mu$m \\
$M_\mathrm{gas}$ & $0.13\ M_\odot$ &
$C_\mathrm{s}$ & $C_0\,r^{-3/8}$, $C_0=297.9$ m\,s$^{-1}$ \\
$M_\mathrm{dust}$ &$0.01\ M_\mathrm{gas}$ & $\Sigma$ & $\Sigma_0\,r^{-3/2}$ \\
$R_\mathrm{disk}$ & 100 to 400 AU & $\alpha_\mathrm{SPH}$ & $\zeta=0.1$ \\
$\rho_\mathrm{d}$ & 1.0 g\,cm$^{-3}$ & $\beta_\mathrm{SPH}$ & 0.0 \\
\hline
\end{tabular}
\end{table}

We run hydrodynamical simulations for a circumbinary disk around two
stars of 0.65 and 0.5~$M_\odot$ respectively, separated by 35~AU and
evolving on orbits with 0.25 eccentricity \citep{Dutrey94}.  We start
from a disk with a 0.5~AU central hole and let the binary dig the hole
during the $10\,000$ first time-steps ($\simeq$ 8 orbits at 100~AU)
while the gas relaxes to a quasi equilibrium state before dust is
injected. We then let the whole evolve for $40\,000$ more time-steps
($\simeq$ 32 orbits at 100~AU).

The disk mass is 0.13~$M_\odot$ with 1\% dust in mass, its outer
boundary is initially fixed to 350~AU and doesn't significantly spread
beyond 400~AU.  Each simulation includes about $25\,000$ particles
(gas+dust). The grain sizes range from 1 to 10~$\mu$m, we omit larger
grains because their contribution to the scattered light images is
negligible. Table~\ref{tab:paramCB} summarizes the simulation
parameters.

The inner edge radius obtained in the simulations, about a hundred AU,
is appreciably smaller than GG~Tau's (180~AU). This is due to the
uncertainty on the binary parameters for GG~Tau and the depletion of
the disk's central regions is therefore different. This simply
introduces a scaling factor and does not alter in any way the results
presented here. More recent binary parameters  that could better
account for the location of ring's inner edge have been obtained but
still need to be confirmed by further astrometric monitoring, as noted
by \citet{Beust05,Beust06}.

\subsection{Simulation results}
\label{sec:hydro_results}

\begin{figure}
\resizebox{\hsize}{!}{
\includegraphics{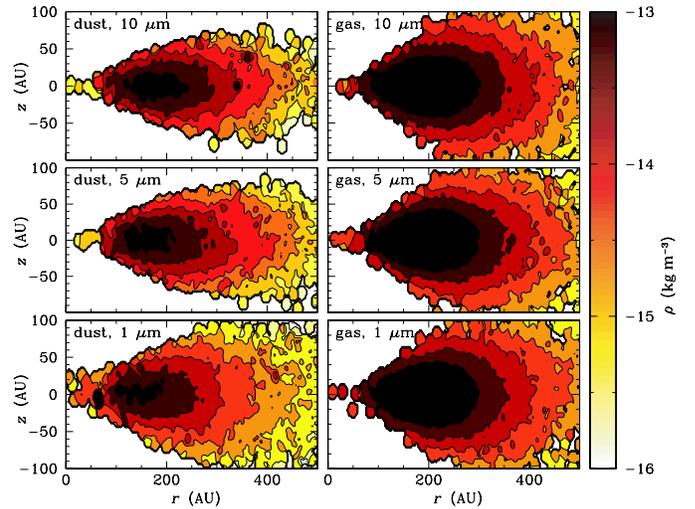}
}
\caption{Density contours for dust (left) and gas (right) at the end
  of the SPH simulations of the circumbinary disk for runs including
  10, 5 and 1~$\mu$m grains from top to bottom. 
 }
\label{GGTau_smsizes}
\end{figure}

Figure~\ref{GGTau_smsizes} displays the dust and gas densities in the
($r$,$z$) plane at the end of the SPH simulations. The settling of
dust appears clearly compared to the gas disk. The outer dust density
contours show it to be more efficient for larger grains (10 $\mu$m)
than for smaller ones (1 $\mu$m) that remain strongly coupled to the
gas.  The radial distribution also differs from one grain size to the
next.  The 1~$\mu$m grains move closer to the star than the 5 and
10~$\mu$m grains (see also Fig.~\ref{fig:Alls}), which provides
physical support to our parametric model of radial and vertical
stratification (Sect.~\ref{sec:modeles_GG}) in which the small grains
form a layer surrounding the central parts of the ring where the large
grains are present.

The reason for this radial segregation is that larger grains respond
more strongly to the pressure gradient and are more efficiently
concentrated in the gas pressure maxima. Therefore, the radial extent
of the density distribution decreases as the grain size increases
(this is more easily seen in Fig.~\ref{fig:Alls}). Any asymmetry
between the inner and outer pressure gradients will produce a shift in
the peak of the density distribution according to the grain size. In
the case of GG~Tau, the pressure gradient is stronger in the inner
side than in the outer side, which results in the density peak being
shifted outwards for larger grains.

\subsection{Interfacing the two codes: analytical fits to the data}
\label{sec:interface}

The SPH data are unfortunately too noisy to be used directly in
MCFOST, even after smoothing on a regular grid. We therefore derive
analytical fits that must consistently reproduce this differential
settling. To do this, we first perform a vertical fit of the dust
density at the end of the SPH simulations at a given radius, the
parameters of this fit are then fitted in the radial direction.

\subsubsection{Vertical fit}
\label{sec:VertFit}

We first decide of a shape for the fit in the vertical direction.
According to \citet{Garaud04b}, the time evolution of the dust density
with height is self-similar, with initial conditions satisfying the
hydrostatic equilibrium. This leads to the classical expression of
density for a geometrically thin disk (in the vertically isothermal
case):

\begin{equation}
\rho(z) = \frac{\Sigma}{\sqrt{2 \pi} H} \exp { \left( -\frac{z^2}{2 H^2} \right) }.
\label{rhoth}
\end{equation}
Because of the self-similarity, the dust density profile remains Gaussian with a width and amplitude that evolve with time.

We produce a vertical fit at a given radius $r_0$ with two parameters $\rho_\mathrm{max}$ and $\sigma_0$ so that
\begin{equation}
\rho(z) = \rho_\mathrm{max}\ \exp \left( \frac{z^2}{2\ \sigma_0^2} \right),
\label{eq:fitz}
\end{equation}
implying that $H = \sigma_0$ and $\Sigma = \sqrt{2 \pi}\ \sigma_0\ \rho_\mathrm{max}$.

\subsubsection{Radial fit}
\label{sec:RadFit}

\begin{figure}
\includegraphics[width=\hsize]{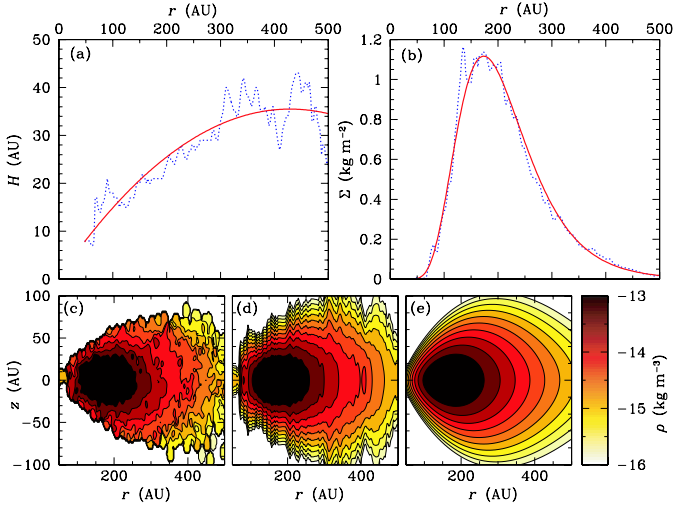}
\caption{Fits for the 5~$\mu$m dust particles in the circumbinary disk.
  \textbf{a.}~Scaleheight. \textbf{b.}~Surface density. In both cases,
  the dotted line corresponds to the vertical fit only and the solid line
  to the subsequent radial fit.
  \textbf{c.}~SPH density after smoothing on a regular grid.
  \textbf{d.}~Vertical fit of the density. \textbf{e.}~Complete fit of the
  density in both vertical and radial directions.}
\label{fig:fitr}
\end{figure}

For the radial fit, we use a parabola to fit the scaleheight $H$
(see Fig.~\ref{fig:fitr}a): 
\begin{equation}
H(r) = a_2 r^2 + a_1 r + a_0.
\end{equation}
We can thus reproduce a flared disk as well as a self shadowed one, as
observed for the dust layer. This behavior of the aspect ratio has a
strong influence on the SED and the synthetic images as is highlighted
by \citet{Dullemond04} and as we will see further on.

In the case of a circumbinary disk, the surface density, plotted
against the logarithm of the radius ($\ln\Sigma = f(r)$), looks like a
Gaussian because the disk is actually a wide ring and there is a
density drop towards the center as well as towards the outside of the
disk.  We choose a fitting formula where we look for the value of the
three parameters $\Sigma_\mathrm{max}$, $r_\mathrm{0max}$, $\sigma'_0$, verifying
\begin{equation}
\Sigma = \Sigma_\mathrm{max} \exp \left( \frac{(\ln(r)-\ln(r_\mathrm{0max}))^2}{2 \sigma^{'2}_0} \right) . 
\end{equation}
The result of this fit is shown on Fig.~\ref{fig:fitr}b for 5~$\mu$m grains.

Finally, with the fits of $H$, of $\Sigma$ and the theoretical
expression of the density (Eq.~\ref{rhoth}), we get an analytical
expression for it:

\begin{eqnarray} 
\rho(r,z) & = & \frac{\Sigma_\mathrm{max}}{\sqrt{2 \pi}\ \sum_{i=0}^2 a_i r^i} \nonumber\\ 
 & &\times \exp { \left( \frac{\left(\ln(r)-ln(r_\mathrm{0max})\right)^2}{2 \sigma^{'2}_0} 
-\frac{z^2}{2 \left( \sum_{i=0}^2 a_i r^i \right)^2} \right) }.
\end{eqnarray} 

This expression of the volumic density gives a much smoother
distribution than the initial data (see Figs.~\ref{fig:fitr}c-e).

\begin{figure}
\includegraphics[width=0.49\hsize]{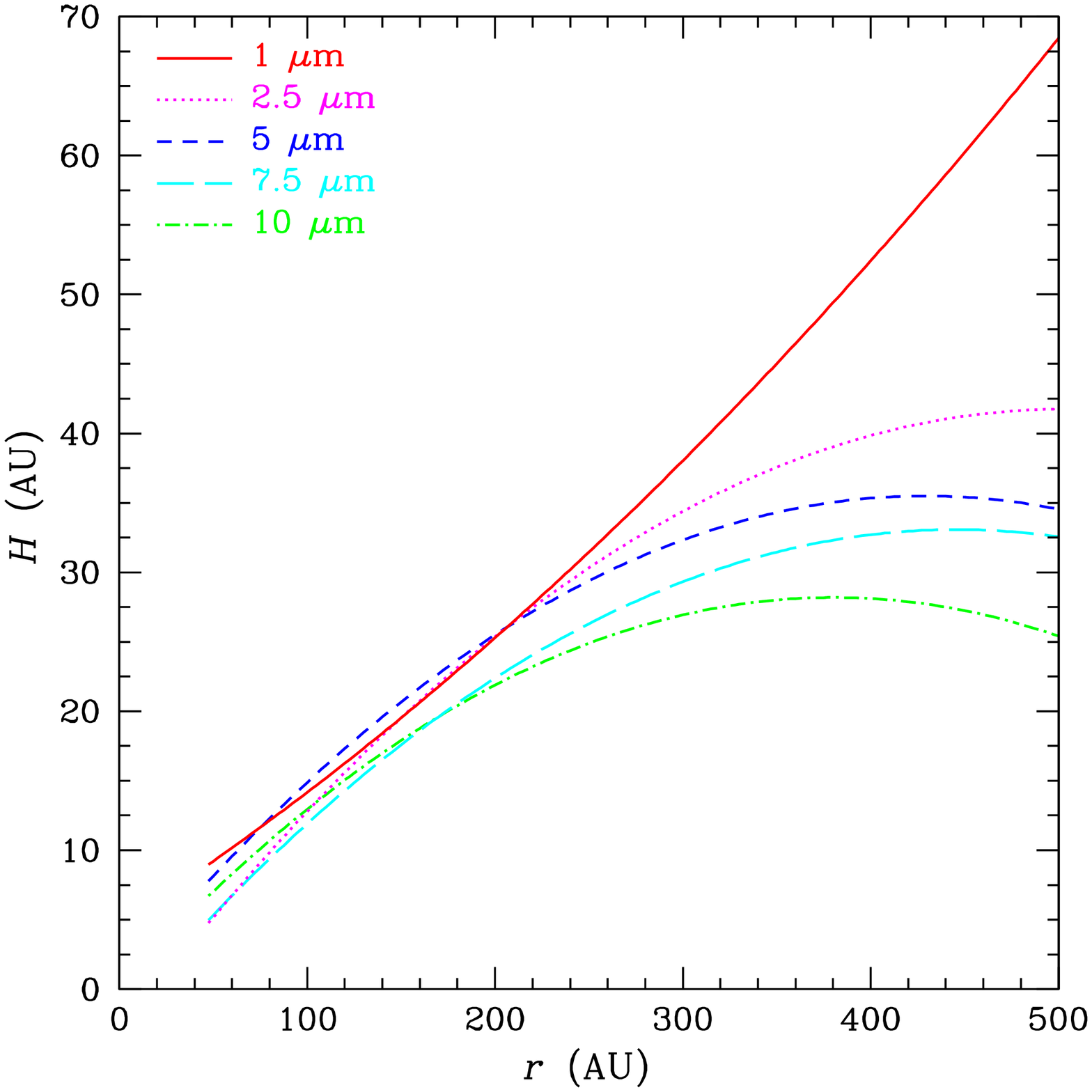}
\includegraphics[width=0.49\hsize]{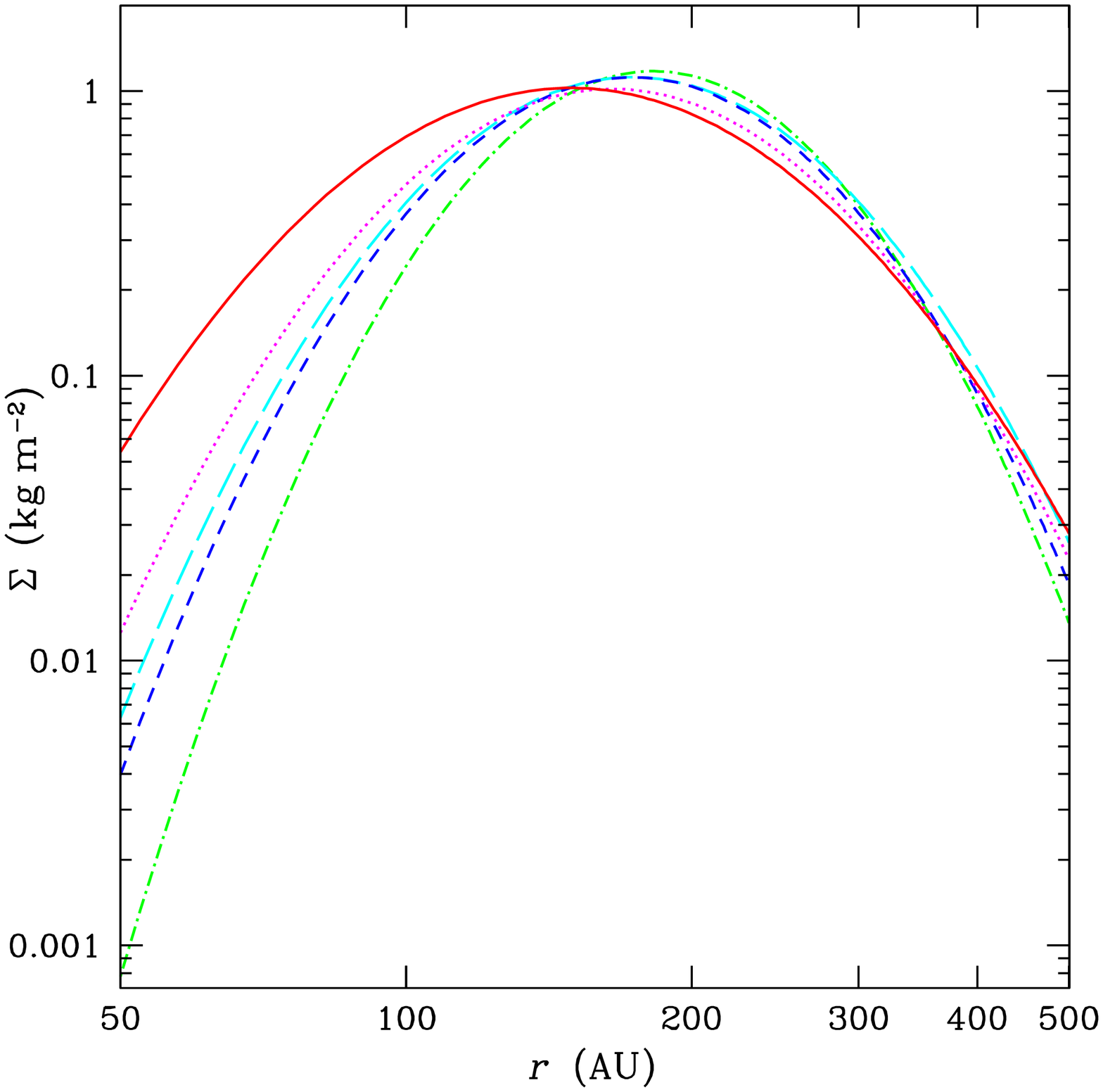}
\caption{Scaleheight (left) and surface density (right) for the small grains (1 to 10 $\mu$m) in the circumbinary disk.
}
\label{fig:Alls}
\end{figure}

The fitted scaleheights and surface densities for different grain
sizes are plotted in Fig.~\ref{fig:Alls}, which clearly shows the
differential settling, on the left panel, with the larger grains
having the more important settling (because we are in the small
particle regime, see \citet{Barriere05}). The right panel shows that
the dust density maximum shifts towards larger radii as the grain size
increases, and therefore as grains settle more efficiently in this
size range (see Sect.~\ref{sec:hydro_results}).

\subsubsection{Interpolation between grain sizes}
\label{sec:Ints}

Each hydrodynamical simulation only includes one grain size at a
time. To synthesize scattered light images, we need to reconstruct a
grain size distribution at each point of the disk. Because we do not
consider interactions between dust grains and because all simulations,
whatever the grain size used, predict consistent density for the gas
phase, we can do this by superimposing the calculated densities for
each grain size. The relative fraction of grains is fixed by assuming
a grain size distribution, integrated over the disk, described as a
power law with an index of $-3.5$.  Finally, an interpolation,
logarithmic with respect to grain sizes, is performed between the
spatial distributions of grains resulting from the SPH calculations,
to get a well sampled grain size distribution. The smallest grains
($a_\mathrm{min} = 0.03\,\mu$m) are assumed to follow perfectly the
gas distribution.

\subsection{Synthetic images}
\label{sec:CBres}

\begin{figure}[t]
  \centering
  \psfrag{intensite normalisee}[c][c]{Normalized intensity}
  \psfrag{angle}[c][c]{Position angle ($^\circ$)} 
  \psfrag{Bande I}[c][c]{I band}
  \psfrag{Bande J}[c][c]{J band}
  \psfrag{Bande H}[c][c]{H band}
  \psfrag{Bande L'}[c][c]{L' band}
  \includegraphics[width=0.9\hsize]{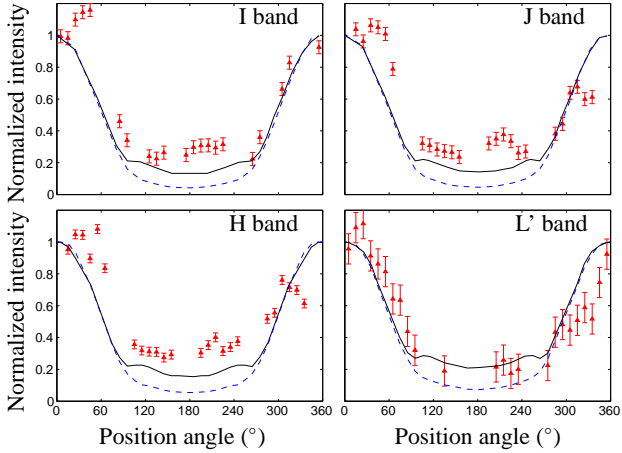}
  \caption{Azimuthal brightness profiles of the circumbinary disk with
    (solid line) and without (dashed line) stratification superimposed to
    GG~Tau's observed profiles (triangles). \label{fig:profil_gg_strat}}
\end{figure}

Figure~\ref{fig:profil_gg_strat}
compares the azimuthal brightness profiles
in the I, J, H and L' bands resulting from our simulations of the
circumbinary disk, with and without grain stratification. In the
latter case, the aerodynamic drag of gas on dust is switched off and
the dust distribution follows that of the gas for all grain sizes. The
differences between models with and without stratification are
especially apparent on the back side of the disk. As for the analytic
stratification we used in Sect.\ref{sec:modeles_GG_strat}, we observe
the smaller grains at the inner rim of the disk. They are isotropic
scatterers and send back a larger fraction of light than the larger
grains, of which we detect the scattered light when all grains are
well mixed.
We want to stress here that
the models are \emph{not} fits to the data, which are plotted as a
reference in order to study the \emph{qualitative} behaviour of our models.

The inclusion of dust segregation is a step in the right direction to
explain GG~Tau's observations. In particular, we note that the back
side of the disk is particularly sensitive to the radial
stratification. The light coming from the front
surface of the ring, which corresponds to angles between 0 and
90$^\circ$ on one side and 270 and 360$^\circ$ on the other side, is
in fact little affected by the segregation and both brightness
profiles (with and without stratification) are almost
identical. However, the back side is noticeably brighter in the
stratified case, for the reasons mentioned above.

It appears that the stratification resulting from the present
hydrodynamical model is not sufficient to explain the observed
brightness profiles.  This may be due to the relatively short time on
which the dynamical evolution is followed, or to physical processes
which also lead to a spatial segregation of grains in the disk but are
not considered here, like a preferred grain growth in the denser parts
of the disk. Nonetheless, our hydrodynamical simulations demonstrate that
it is physically plausible that the ring possesses an ``inner wall'' of
small grains, as suggested by our parametric approach.

\section{Discussion}
\label{sec:discussion}

The inclusion of stratification in a parametric model of GG~Tau's disk
(Sect.~\ref{sec:modeles_GG}) allows to reproduce the scattered light
images from the I to the L' band, in particular the azimuthal
brightness distribution which is remarkably similar at all
wavelengths.  This model is also in good agreement with the 1~$\mu$m
polarisation data and globally reproduces the ring colours with
respect to the central star.  However, a simple parametric vertical
stratification alone cannot explain all observations, especially the
brightness of the inner rim. The presence of a radial segregation
at the inner edge, with smaller grains closer to the stars than larger
grains, also appears required to obtain synthetic images which
reproduce the observations. This kind of distribution can result from
a preferred grain growth in the central, denser zones of disks where
the collision probability is higher. It is also predicted by
simulations of the dust dynamics around a binary system (see
Sect.~\ref{sec:hydro_results}), which presents synthetic scattered light
images showing a good qualitative match with the observations, as
shown in Sect.~\ref{sec:CBres}.

Adding dust stratification and settling into models that already
contain several free parameters renders the problem of finding unique
solutions even more acute. We explored a large parameter space and
tried to quantify, through Bayesian analyses, the likelihood of
occurrence of the models we presented. By doing so, we also found a
solution {\sl with well-mixed dust} that also matched the observed
azimuthal profiles, at all wavelengths. However, this solution appears
physically less realistic than the expected one with settling because
it requires a very steep slope for the dust size distribution.  Yet,
the solution exists and we added observational constraints to confirm
or reject its validity.  When compared to polarisation data of the
ring, the ``well-mixed'' solution proves only marginally compatible
with the data, and certainly worse at matching observations than the
model with stratification. More interesting, the well-mixed models
we explored are not able to reproduce the correct dependency of disk
colours with respect to the star as a function of wavelength.

The best ``well-mixed'' solution we identified is made of a
distribution of spherical and homogeneous particles, with a slope of
$-5.5$, noticeably steeper than that of interstellar medium
grains. 
Models of dust coagulation \citep{Dullemond05,Ormel07} show that, at
least during the first steps of the process, the slope of the grain
size distribution does not vary so much, in the size regime we are
interested here. We succeeded until now in fitting a growing number of
observable quantities by modifying the grain properties in a
well-mixed model but this fit is still not perfect, and noticeably
poorer than in the stratified case. These modifications are
increasingly arbitrary and it becomes difficult to find physical
causes to explain them.  Therefore, we consider this solution with a
steep dust size distribution more ``exotic'' and less likely to
occur. The model with stratification provides a better match to more
data sets. Furthermore, stratification appears a more natural outcome
of the disk evolution processes and for now it is our preferred
model for the ring of GG~Tau.

\cite{Krist05} show evidence that the azimuthal brightness
  profiles change with time with front/back amplitude variations
  around 20\,\%. These fluctuations remain moderate and taking them
  into account only slightly affects the major observational property
  of the ring in scattered light : the similar
  morphology of the disk over a large range of wavelengths.
 Therefore, our general conclusion that dust stratification is required to
 interpret scattered light images still holds but a more precise
 determination of the degree of settling will probably require
 comtemporaneous multiple wavelengths data sets.

In all cases, the presence of grains larger than a few microns is
required.  Both models require a minimum value of $a_\mathrm{max}$ of
the order of 3~$\mu$m, indicating that the grains have evolved
compared to those of the interstellar medium. The value inferred from
our simulations assuming spherical grains is indeed a lower limit, a
particle of complex shape being able to mimic the scattering
properties of a smaller spherical and homogeneous particle \citep{Rouleau96}. This implies that grain growth towards sizes larger
than one micron is happening in this circumbinary ring and that the
first steps of planet formation are very probably being initiated.

\section{Conclusions}
\label{sec:conclusion}

The definitive proof of a stratified structure in GG~Tau's disk will
probably rely on additional observations. However, stratified dust
models are able to match several observable quantities of the GG~Tau
ring: in particular the azimuthal brightness profiles from 800~nm
to 3800~nm, but also polarisation measurements at 1~micron and
the variation of star/disk flux ratios with wavelength.

We have shown the attractiveness of polarimetric observations to
better constrain the grain properties. 
The broadening of the wavelength coverage also
seems very promising, both in intensity and polarimetry. The detection
of the disk at 10~$\mu$m, where our models show that scattered light
still dominates over the disk thermal emission, would enlarge the
wavelength leverage at our disposal and probe layers even closer to
the disk midplane than the current 25~AU\footnote{Depending on grain
models, and in particular on the fraction of silicates in the grains,
the relative opacities at 3.8 and 10~$\mu$m vary widely and the zone
that is probed at 10~$\mu$m is difficult to estimate beforehand. On
the other hand, a geometrical study, similar to the one realized by
\D04, at a wavelength close to 10~$\mu$m would give essential
information on this opacity ratio and, as a consequence, on the
composition of grains.}.
The inclusion of the thermal emission allows additionally to derive an
estimate of the ratio between the absorption and scattering
cross-sections and to further restrict the families of solutions
compatible with the observations of a given disk.

The modeling of the circumbinary disk in Sect.~\ref{sec:modeles_hydro}
allowed us to give a physical basis to the parametric modeling of
GG~Tau's ring in Sect.~\ref{sec:modeles_GG}. Hydrodynamic models
validate our hypothesis of a radial stratification of grains and are
in qualitative agreement with the observations. It is indeed to the
grains radial segregation, rather than to their vertical
stratification, that we are  most sensitive in GG~Tau's disk.
The configuration we used does not exactly correspond to GG~Tau's, due
to uncertainties in the binary's orbital parameters. The use of better
parameters should allow us to make more quantitative comparisons with
the data and to start to better constrain the processes causing the
stratification in GG~Tau's disk.

We would like to stress here that validating the parametric approach
by hydrodynamic computations is important because it  allows the
use of a faster tool to quickly estimate whether settling has occurred
in a disk or not, before embarking on dedicated hydrodynamic
calculations, which are more realistic but necessarily longer.

On the long term, it will be necessary to include new physical processes
acting in addition to dust dynamics. Turbulence, in the calculations
presented here, was only taken into account in a simplified manner. It
would be interesting to see how a better treatment of turbulence limits
the settling and affects the observable quantities.

Grain growth
is intimately linked to their settling. Because it enlarges the relative
velocities, settling increases the number of collisions, which favours
grain growth, which in turns increases the settling. The effects on the
different observable quantities are potentially enhanced. The
simultaneous modeling of both processes appears to be the next step towards
a better understanding of the first stages of planet formation. It is
currently being implemented into the SPH code \citep{Laibe07} and the
resulting simulations will be coupled to the radiative transfer code to
continue our study.

\begin{acknowledgements}

Computations presented in this paper were performed at the Service
Commun de Calcul Intensif de l'Observatoire de Grenoble (SCCI)
and at the Centre Informatique National de l'Enseignement Sup\'erieur (CINES).
We thank the {\sl Programme National de Physique Stellaire}
(PNPS) and {\sl l'Action Sp\'ecifique en Simulations Num\'eriques pour
l'Astronomie} (ASSNA) of CNRS/INSU, France, for supporting part of
this research. Finally, we wish to thank the referee, A. Watson, for
his comments which have helped to improve the manuscript.

\end{acknowledgements}

\bibliographystyle{aa}
\bibliography{biblio}

\begin{thebibliography}{38}
\expandafter\ifx\csname natexlab\endcsname\relax\def\natexlab#1{#1}\fi

\bibitem[{{Apai} {et~al.}(2004){Apai}, {Pascucci}, {Sterzik}, {van der Bliek},
  {Bouwman}, {Dullemond}, \& {Henning}}]{Apai04}
{Apai}, D., {Pascucci}, I., {Sterzik}, M.~F., {et~al.} 2004, \aap, 426, L53

\bibitem[{{Barri{\`e}re-Fouchet} {et~al.}(2005){Barri{\`e}re-Fouchet},
  {Gonzalez}, {Murray}, {Humble}, \& {Maddison}}]{Barriere05}
{Barri{\`e}re-Fouchet}, L., {Gonzalez}, J.-F., {Murray}, J.~R., {Humble},
  R.~J., \& {Maddison}, S.~T. 2005, \aap, 443, 185

\bibitem[{{Beust} \& {Dutrey}(2005)}]{Beust05}
{Beust}, H. \& {Dutrey}, A. 2005, \aap, 439, 585

\bibitem[{{Beust} \& {Dutrey}(2006)}]{Beust06}
{Beust}, H. \& {Dutrey}, A. 2006, \aap, 446, 137

\bibitem[{{Bicknell}(1991)}]{Bicknell91}
{Bicknell}, G.~V. 1991, SIAM J. Sci. Atat. Comput., 12, 1198

\bibitem[{{Blum} {et~al.}(2000){Blum}, {Wurm}, {Kempf}, {Poppe}, {Klahr},
  {Kozasa}, {Rott}, {Henning}, {Dorschner}, {Schr{\"a}pler}, {Keller},
  {Markiewicz}, {Mann}, {Gustafson}, {Giovane}, {Neuhaus}, {Fechtig},
  {Gr{\"u}n}, {Feuerbacher}, {Kochan}, {Ratke}, {El Goresy}, {Morfill},
  {Weidenschilling}, {Schwehm}, {Metzler}, \& {Ip}}]{Blum00}
{Blum}, J., {Wurm}, G., {Kempf}, S., {et~al.} 2000, Physical Review Letters,
  85, 2426

\bibitem[{{Dominik} {et~al.}(2006){Dominik}, {Blum}, {Cuzzi}, \&
  {Wurm}}]{Dominik06PPV}
{Dominik}, C., {Blum}, J., {Cuzzi}, J., \& {Wurm}, G. 2006

\bibitem[{{Dubrulle} {et~al.}(1995){Dubrulle}, {Morfill}, \&
  {Sterzik}}]{Dubrulle95}
{Dubrulle}, B., {Morfill}, G., \& {Sterzik}, M. 1995, Icarus, 114, 237

\bibitem[{{Duch{\^ e}ne} {et~al.}(2004){Duch{\^ e}ne}, {McCabe}, {Ghez}, \&
  {Macintosh}}]{Duchene04}
{Duch{\^ e}ne}, G., {McCabe}, C., {Ghez}, A.~M., \& {Macintosh}, B.~A. 2004,
  \apj, 606, 969

\bibitem[{{Dullemond} \& {Dominik}(2004)}]{Dullemond04}
{Dullemond}, C.~P. \& {Dominik}, C. 2004, \aap, 421, 1075

\bibitem[{{Dullemond} \& {Dominik}(2005)}]{Dullemond05}
{Dullemond}, C.~P. \& {Dominik}, C. 2005, \aap, 434, 971

\bibitem[{{Dutrey} {et~al.}(1994){Dutrey}, {Guilloteau}, \& {Simon}}]{Dutrey94}
{Dutrey}, A., {Guilloteau}, S., \& {Simon}, M. 1994, \aap, 286, 149

\bibitem[{{Furlan} {et~al.}(2005){Furlan}, {Calvet}, {D'Alessio}, {Hartmann},
  {Forrest}, {Watson}, {Uchida}, {Sargent}, {Green}, \& {Herter}}]{Furlan05}
{Furlan}, E., {Calvet}, N., {D'Alessio}, P., {et~al.} 2005, \apjl, 628, L65

\bibitem[{{Garaud} \& {Lin}(2004)}]{Garaud04b}
{Garaud}, P. \& {Lin}, D.~N.~C. 2004, \apj, 608, 1050

\bibitem[{{Goldreich} \& {Ward}(1973)}]{Goldreich73}
{Goldreich}, P. \& {Ward}, W.~R. 1973, \apj, 183, 1051

\bibitem[{{Guilloteau} {et~al.}(1999){Guilloteau}, {Dutrey}, \&
  {Simon}}]{Guilloteau99}
{Guilloteau}, S., {Dutrey}, A., \& {Simon}, M. 1999, \aap, 348, 570

\bibitem[{{Jones} {et~al.}(1996){Jones}, {Tielens}, \& {Hollenbach}}]{Jones96}
{Jones}, A.~P., {Tielens}, A.~G.~G.~M., \& {Hollenbach}, D.~J. 1996, \apj, 469,
  740

\bibitem[{{Krist} {et~al.}(2002){Krist}, {Clampin}, {Golimowski}, {Ardila},
  {Bartko}, {Ben{\'{\i}} tez}, {Blakeslee}, {Bouwens}, {Broadhurst}, {Brown},
  {Burrows}, {Cheng}, {Cross}, {Feldman}, {Ford}, {Franx}, {Gronwall},
  {Hartig}, {Illingworth}, {Infante}, {Kimble}, {Lesser}, {Martel},
  {Menanteau}, {Miley}, {Postman}, {Rosati}, {Sirianni}, {Sparks}, {Tran},
  {Tsvetanov}, {White}, \& {Zheng}}]{Krist02}
{Krist}, J.~E., {Clampin}, M., {Golimowski}, D.~A., {et~al.} 2002, Bulletin of
  the American Astronomical Society, 34, 1319

\bibitem[{{Krist} {et~al.}(2005){Krist}, {Stapelfeldt}, {Golimowski}, {Ardila},
  {Clampin}, {Martel}, {Ford}, {Illingworth}, \& {Hartig}}]{Krist05}
{Krist}, J.~E., {Stapelfeldt}, K.~R., {Golimowski}, D.~A., {et~al.} 2005, \aj,
  130, 2778

\bibitem[{{Laibe} {et~al.}(2007){Laibe}, {Gonzalez}, \& {Fouchet}}]{Laibe07}
{Laibe}, G., {Gonzalez}, J.-F., \& {Fouchet}, L. 2007, \aap, in preparation

\bibitem[{{Lay} {et~al.}(1997){Lay}, {Carlstrom}, \& {Hills}}]{Lay97}
{Lay}, O.~P., {Carlstrom}, J.~E., \& {Hills}, R.~E. 1997, \apj, 489, 917

\bibitem[{{Mathis} {et~al.}(1977){Mathis}, {Rumpl}, \& {Nordsieck}}]{Mathis77}
{Mathis}, J.~S., {Rumpl}, W., \& {Nordsieck}, K.~H. 1977, \apj, 217, 425

\bibitem[{{Mathis} \& {Whiffen}(1989)}]{Mathis89}
{Mathis}, J.~S. \& {Whiffen}, G. 1989, \apj, 341, 808

\bibitem[{{McCabe} {et~al.}(2002){McCabe}, {Duch{\^e}ne}, \& {Ghez}}]{McCabe02}
{McCabe}, C., {Duch{\^e}ne}, G., \& {Ghez}, A.~M. 2002, \apj, 575, 974

\bibitem[{{Monaghan}(1992)}]{Monaghan92}
{Monaghan}, J.~J. 1992, \araa, 30, 543

\bibitem[{{Monaghan}(1997)}]{Monaghan97}
{Monaghan}, J.~J. 1997, Journal of Computational Physics, 136, 298

\bibitem[{{Monaghan} \& {Kocharyan}(1995)}]{Monaghan95}
{Monaghan}, J.~J. \& {Kocharyan}, A. 1995, Computer Physics Communications, 87,
  225

\bibitem[{{Murray}(1996)}]{Murray96}
{Murray}, J.~R. 1996, \mnras, 279, 402

\bibitem[{{Ormel} {et~al.}(2007){Ormel}, {Spaans}, \& {Tielens}}]{Ormel07}
{Ormel}, C.~W., {Spaans}, M., \& {Tielens}, A.~G.~G.~M. 2007, \aap, 461, 215

\bibitem[{{Pinte} {et~al.}(2006){Pinte}, {M{\'e}nard}, {Duch{\^e}ne}, \&
  {Bastien}}]{Pinte06}
{Pinte}, C., {M{\'e}nard}, F., {Duch{\^e}ne}, G., \& {Bastien}, P. 2006, \aap,
  459, 797

\bibitem[{{Press} {et~al.}(1992){Press}, {Teukolsky}, {Vetterling}, \&
  {Flannery}}]{Press92}
{Press}, W.~H., {Teukolsky}, S.~A., {Vetterling}, W.~T., \& {Flannery}, B.~P.
  1992, {Numerical recipes. The art of scientific computing} (Cambridge:
  University Press, |c1992, 2nd ed.)

\bibitem[{{Rouleau}(1996)}]{Rouleau96}
{Rouleau}, F. 1996, \aap, 310, 686

\bibitem[{{Safronov} \& {Zvjagina}(1969)}]{Safronov69}
{Safronov}, V.~S. \& {Zvjagina}, E.~V. 1969, Icarus, 10, 109

\bibitem[{{Silber} {et~al.}(2000){Silber}, {Gledhill}, {Duch{\^e}ne}, \&
  {M{\'e}nard}}]{Silber00}
{Silber}, J., {Gledhill}, T., {Duch{\^e}ne}, G., \& {M{\'e}nard}, F. 2000,
  \apjl, 536, L89

\bibitem[{{Stepinski} \& {Valageas}(1996)}]{Stepinski96}
{Stepinski}, T.~F. \& {Valageas}, P. 1996, \aap, 309, 301

\bibitem[{{Watson} {et~al.}(2006){Watson}, {Stapelfeldt}, {Wood}, \&
  {M\'enard}}]{Watson06PPV}
{Watson}, A., {Stapelfeldt}, K., {Wood}, K., \& {M\'enard}, F. 2006

\bibitem[{{Weidenschilling}(1977)}]{Weidenschilling77}
{Weidenschilling}, S.~J. 1977, \mnras, 180, 57

\bibitem[{{Wood} {et~al.}(2002){Wood}, {Wolff}, {Bjorkman}, \&
  {Whitney}}]{Wood02}
{Wood}, K., {Wolff}, M.~J., {Bjorkman}, J.~E., \& {Whitney}, B. 2002, \apj,
  564, 887

\end{thebibliography}
\end{document}